%% file: Rogochaya_WPCF_proceedings.tex
\documentclass[WPCF,manyauthors]{wpcfTemplate}

\usepackage{rotating}
\usepackage{subfigure}

\usepackage{lineno}
\usepackage{hyperref}
\begin{document}%
%%%%%%%%%%%%% ptdr definitions %%%%%%%%%%%%%%%%%%%%%
%
%%%%%%%%%%%%%%%  Title page %%%%%%%%%%%%%%%%%%%%%%%%
%
{\let\cleardoublepage\clearpage
\begin{titlepage}
%
%%%
\title{Charged kaon femtoscopic correlations \\ in p--Pb
collisions at $\mathbf{\sqrt{{\textit s}_{\rm NN}}}=5.02$~TeV  with ALICE at the LHC}
\ShortTitle{Charged kaon femtoscopic correlations}   % appears on right page headers
\author{E.P. Rogochaya$^{1}$ (for the ALICE Collaboration)}
\author{
1. Joint Institute for Nuclear Research, Joliot-Curie 6, 141980 Dubna, Moscow region,
Russia \\
}
\author{Email: elena.rogochaya@cern.ch}
\ShortAuthor{XII$^{th}$ edition of WPCF}      % appears on left page headers, do not change
\begin{abstract}
Particle correlations at small relative momenta can be used to measure the space--time characteristics of particle production on the femtoscopic (fm=10$^{-15}$~m) level in high-energy collisions. Such correlations arise due to quantum statistics and final state interactions. We report correlations of two charged identical kaons measured in p--Pb collisions at $\sqrt{s_{\rm NN}}=5.02$~TeV by the ALICE experiment at the LHC. The femtoscopic invariant radii and correlation strengths are extracted from the one-dimensional kaon correlation functions and are compared to those obtained in pp and Pb--Pb collisions at $\sqrt{s}=7$~TeV and $\sqrt{s_{\rm NN}}=2.76$~TeV, respectively. Kaon femtoscopy in p--Pb collisions is an important supplement to that in pp and Pb--Pb collisions because it allows one to understand the particle production mechanisms in different collision systems. It also complements the existing pion correlation results. The obtained radii increase with increasing multiplicity and decrease with increasing pair transverse momentum. At comparable multiplicity, the radii measured in p--Pb collisions are close to those observed in pp collisions, whereas it is difficult to compare them with the results from Pb--Pb collisions because of a big multiplicity gap.
\end{abstract}
\end{titlepage}
}
\input{main.tex}               %%%%%%%%%%% put the body of the article here

\input{biblio.tex}
%\addcontentsline{toc}{section}{\bibname}
%\bibliography{biblio}

\end{document}

%% file: main.tex
\section{Introduction}
%These proceedings consider the Bose--Einstein correlations of
%identical charged kaons in p--Pb collisions at
%$\sqrt{s_{\mathrm{NN}}}=5.02$~TeV performed by the ALICE
%collaboration at the LHC from data taken in 2013.
The Bose--Einstein
enhancement of the production of two identical pions at low relative
momenta was first observed in $\bar{\rm p}$p collisions about 50
years ago \cite{Goldhaber:1960sf}. Since that time the correlation
method has been developed and now it is known as ``correlation
femtoscopy'' \cite{Kopylov:1972qw,Kopylov:1974uc}. Femtoscopy measures the apparent width of the
distribution of the relative separation of emission points, which
is conventionally called the ``radius parameter''. The method was successfully applied to the measurement of
the space--time characteristics of particle production processes at
high energies in particle \cite{Kittel:2001zw,Alexander:2003ug}
and heavy-ion collisions (see, e.g.,
\cite{Podgoretsky:1989bp,Lednicky:2003mq} and references therein).

The ALICE collaboration has already studied pion correlations in p--Pb collisions at
$\sqrt{s_{\mathrm{NN}}}=5.02$~TeV \cite{Adam:2015pya}. Extensive ALICE PID capabilities \cite{Aamodt:2008zz} and availability of a large amount of experimental data allow performing the
K$^{\pm}$K$^{\pm}$ femtoscopic analysis in p--Pb collisions at the same energy, which is the aim of this work.
Kaons are a convenient tool to study, in particular, Bose--Einstein correlations. They are less influenced by resonance decays than pions and therefore more effectively probe femtoscopic correlations of directly-produced particles.
%Kaons are less influenced by resonance decays than pions. So, the kaon femtoscopic correlation signal is clearer or easier to distinguish from non-femtoscopic signals and therefore easier to study.
The comparison of kaon and pion correlation radii \cite{Adam:2015pya,Abelev:2014pja} depending on pair transverse momentum $k_{\rm T}=
|{\bf p}_{\mathrm{T,1}} + {\bf p}_{\mathrm{T,2}}|/2$ or
transverse mass $m_{\rm T}= \sqrt{<k_\mathrm{
T}>^2+m^{2}}$, where $m$ is a correlating particle mass, allows one to understand collective dynamics (collective flow) of a source created in collisions. In addition, the kaon radii in p--Pb collisions at $\sqrt{s_{\mathrm{NN}}}=5.02$~TeV are compared with those in pp and Pb--Pb collisions at $\sqrt{s}=7$~TeV \cite{Abelev:2012sq} and $\sqrt{s_{\mathrm{NN}}}=2.76$~TeV \cite{Adam:2015vja}, respectively. It provides experimental constraints on the validity of hydrodynamic \cite{Bozek:2013df} and/or color glass condensate \cite{Bzdak:2013zma} approaches
%(collective radial flow, effects related to the uncertainty principle, initial-state shape and size)
proposed for interpretation of the p--Pb data. In heavy-ion collisions, the decrease of the correlation radii
with increasing $k_{\mathrm T}$ ($m_{\mathrm T}$) was usually considered as a manifestation of
collective behavior of matter created in such collisions. If the
dependence of the interferometry radii on pair momentum in p--Pb
collisions follows the trends seen in heavy-ion collisions, it
would be an indication of collectivity \cite{Bozek:2013df}.

\section{Analysis}
%The data used in the analysis come from p--Pb collisions at 5.02~TeV recorded by ALICE during the 2013 run at the LHC.
The data sample used in the analysis was recorded with the ALICE detector \cite{Aamodt:2008zz} in 2013 at a center-of-mass energy per
nucleon of $\sqrt{s_{\mathrm{NN}}}=5.02$~TeV.
The analysis was performed in three multiplicity bins:
0--20\%, 20--40\%, 40--90\% and two pair transverse momentum $k_{\mathrm T}$
bins: (0.2--0.5), (0.5--1.0)~GeV/$c$. The multiplicity classes are determined \cite{Adam:2014qja} with the V0 detector \cite{Abbas:2013taa}. The identification of kaons was performed using the Time Projection Chamber (TPC) detector for all particle momenta and the Time-Of-Flight (TOF) detector -- for $p>0.5$~GeV/$c$. To ensure uniform tracking, events with the collision vertex position within $\pm$10~cm from the center of TPC, measured along the beam axis $z$, are selected. At least one particle in the event has to be reconstructed and identified as a charged kaon. The correlation signal is constructed from events having at least two same-charged kaons. The one-kaon events contribute to the background determination.
%The transverse momentum of the TPC tracks used in this analysis is required to be 0.14$<p_{\mathrm T}<$1.5~GeV/$c$. The magnitude of a track's pseudorapidity is required to be less than 0.8.
The charged tracks used in this analysis are required to have a transverse momentum 0.14$<p_{\mathrm T}<$1.5~GeV/$c$ and a pseudorapidity $|\eta|< $0.8.
Usually the femtoscopic correlation functions (CF) of identical particles are very sensitive to the two-track reconstruction effects because the considered particles have close momenta and close trajectories. Two kinds of two-track effects were investigated. The track ``splitting'' occurs when one track is mistakenly reconstructed as two. The ``merging'' is the effect when two different tracks are reconstructed as one. To decrease the influence of these effects two-track cuts are applied on the pseudorapidity $\Delta\eta$ and the modified azimuthal angle $\Delta\varphi^*$ that takes into account the bending inside the barrel magnetic field and is measured at the radial distance of 1.6~m.

%The momentum correlations are usually studied using correlations of two or more particles.
The two-particle correlation function $C({\bf p}_1,{\bf p}_2)=A({\bf
p}_1,{\bf p}_2)/B({\bf p}_1,{\bf p}_2)$ is defined as a ratio of
the two-particle distribution in the given event $A({\bf p}_1,{\bf
p}_2)$ to the reference one $B({\bf p}_1,{\bf p}_2)$ \cite{Kopylov:1974th}. Here the
reference distribution is constructed by mixing particles from
different events of a given class.

The correlation function is measured as a function of the invariant pair relative momentum
$q_\mathrm{inv}= \sqrt{|{\bf q}|^{2} - q_{0}^{2}} $, where $q_0=E_1-E_2$ and ${\bf q}={\bf p_1}-{\bf p_2}$ are determined by the energy components $E_1$, $E_2$ and momenta ${\bf p_1}$, ${\bf p_2}$ of particles, respectively.
%${\bf q}={\bf p_1}-{\bf p_2}$ is determined by the momenta of particles ${\bf p_1}$, ${\bf p_2}$ and $q_0=E_1-E_2$ -- by their energy components $E_1$, $E_2$.
The CFs can be parametrized by various formulae depending on the origin of correlations between considered particles. Identical charged kaons correlations are induced by quantum statistics and Coulomb interaction effects. Strong final-state interactions between kaons are negligible \cite{Beane:2007uh}.
Assuming a Gaussian distribution of a particle source
in the pair rest frame, the fitting of the kaon CF is performed using the Bowler--Sinyukov formula \cite{Sinyukov:1998fc,Bowler:1986ta}
\begin{equation}
CF(q_\mathrm{inv})=N\left(1 -\lambda +\lambda K(q_\mathrm{inv})\left( 1+
\exp{\left(-R_\mathrm{inv}^{2} q^{2}_\mathrm{inv}\right)}\right)\right)\,D(q_\mathrm{inv}).
\label{eq:BS_CF}
\end{equation}
The factor $K(q_\mathrm{inv})$ describes the Coulomb interaction, $D(q_\mathrm{inv})$ parametrizes the baseline including all non-femtoscopic
effects, for instance resonance decays, $N$ is a normalization. The Coulomb interaction is determined as
\begin{eqnarray}
K=\frac{C({\rm QS+Coulomb})}{C({\rm QS})},\label{Coulomb}
\end{eqnarray}
where $C({\rm QS})$ is a theoretical CF calculated with pure quantum statistic (QS) weights (wave function squared) and $C({\rm QS+Coulomb})$ corresponds to quantum statistic + Coulomb weights \cite{Sinyukov:1998fc,Abelev:2013pqa}.
The parameters $R_\mathrm{inv}$ and $\lambda$ describe the size of the
source and the correlation strength, respectively.

\section{Results}
\begin{figure}[ht]
\begin{center}
\includegraphics[width=0.7\textwidth]{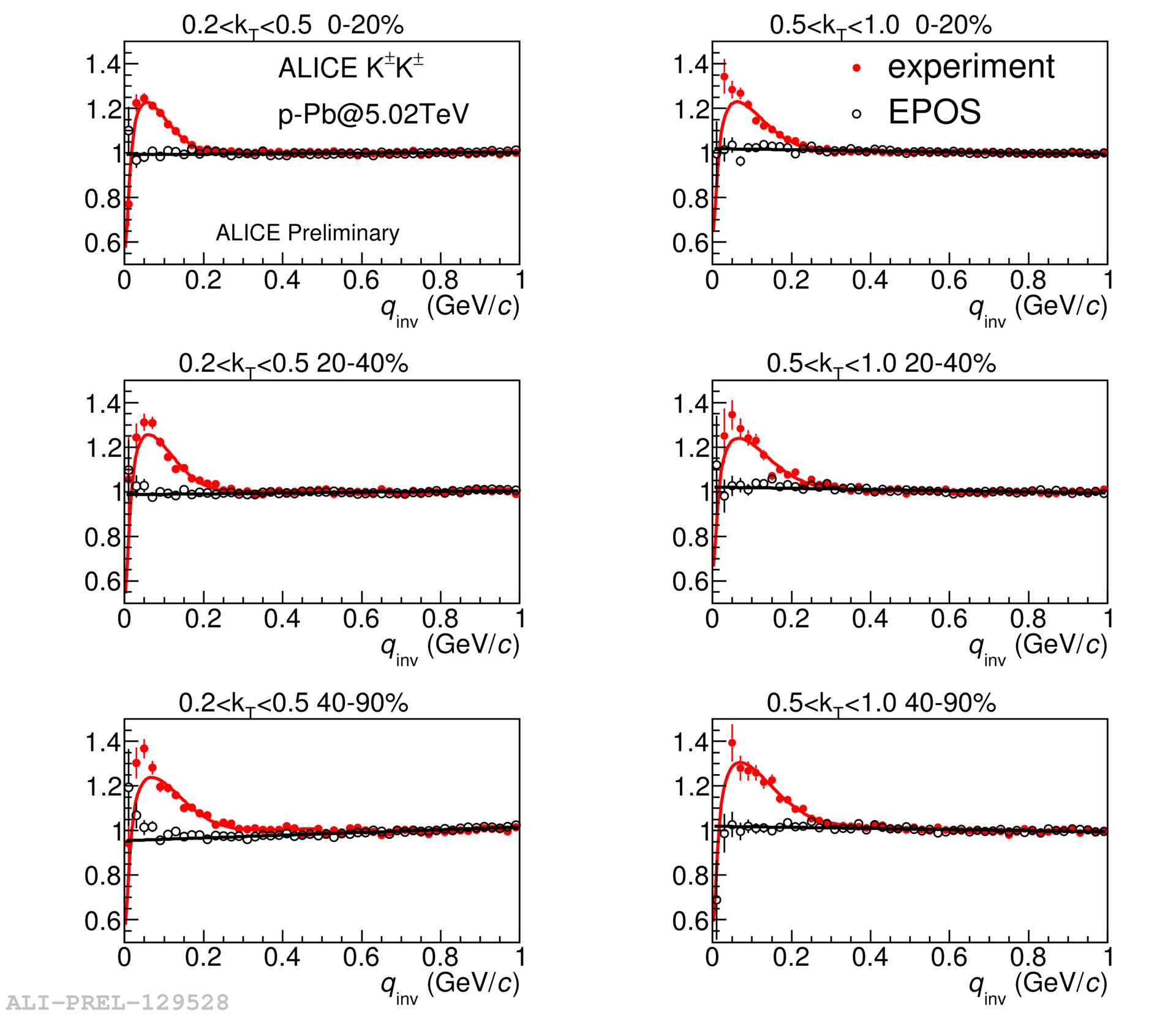}
\caption{K$^\pm$K$^\pm$ experimental correlation functions (red solid markers) versus pair relative invariant momentum $q_\mathrm{inv}$ fitted by (\ref{eq:BS_CF}) up to $q_\mathrm{inv}<$0.5~GeV/$c$ (red line) if baseline is described by the EPOS model \cite{EPOS3} (black empty markers) in the $0<q_\mathrm{inv}<$1.0~GeV/$c$ range. Black line shows fit of EPOS by a first-order polynomial.}
\label{fig:2017-Jun-05-CFs_fit_exp_epos}
\end{center}
\end{figure}

The parameters $R_\mathrm{inv}$ and $\lambda$ can be extracted from (\ref{eq:BS_CF}) using various assumptions %to describe the Coulomb interaction $K$ and
to handle the effects of the baseline $D$.
In this work, the EPOS model \cite{EPOS3} without femtoscopic effects included is used to describe the baseline $D(q_\mathrm{inv})$. First, since it is a realistic model, which reproduces all major results of existing accelerator data \cite{Pierog:2013ria}, and, second, since it gives the best description of kaon spectra \cite{Abelev:2013haa}. Fitting the EPOS baseline by a first-order polynomial in $0<q_{\rm inv}<1.0$~GeV/$c$ and then the experimental CF by (\ref{eq:BS_CF}) in $0<q_{\rm inv}<0.5$~GeV/$c$, one can extract the femtoscopic characteristics. The Coulomb interaction radius is set to $r=1.5$~fm, which is on average close to the extracted radii values. The result is shown in Fig.~\ref{fig:2017-Jun-05-CFs_fit_exp_epos}. The CFs are normalized to unity in the $0.25<q_\mathrm{inv}<0.5$~GeV/$c$ range. As it is seen from the figure, EPOS describes the experimental data outside the femto peak. The experimental CF is flat outside the peak as well as the EPOS function.

\begin{figure}%[htbp]
\centering
\subfigure[~]{\label{fig:RL_sysA}\includegraphics[width=0.49\linewidth]{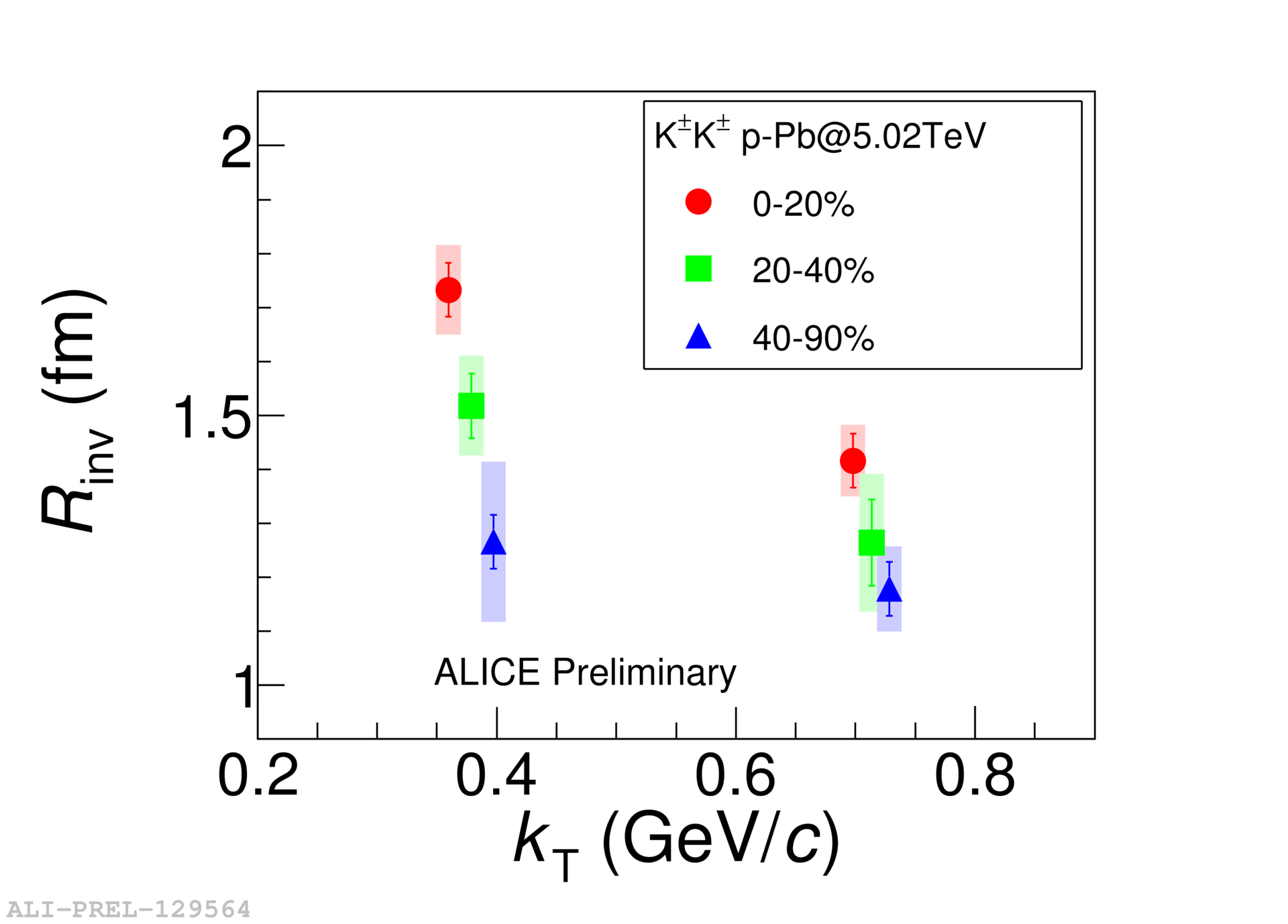}}
\subfigure[~]{\label{fig:RL_sysB}\includegraphics[width=0.49\linewidth]{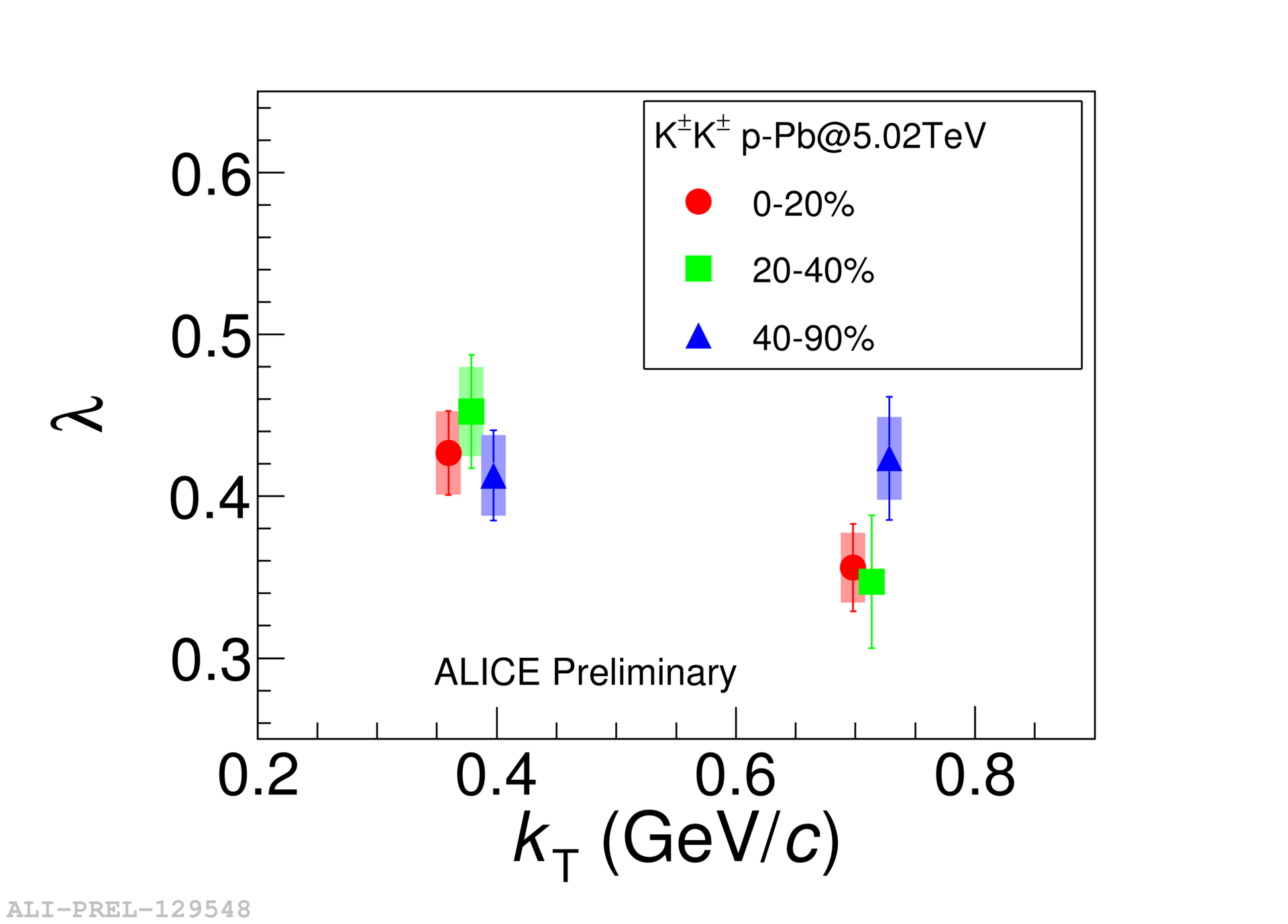}}
\caption{K$^\pm$K$^\pm$ invariant radii
$R_\mathrm{inv}$ (a) and correlation strengths $\lambda$ (b) versus
pair transverse momentum $k_{\mathrm T}$. Statistical errors (lines) and systematic uncertainties (full rectangles) are shown. The points for different centralities
are slightly offset in the $x$ direction for clarity.} %\label{fig:RL_sys}
\end{figure}

\begin{figure}[h]
\begin{center}
\includegraphics[width=0.5\textwidth]{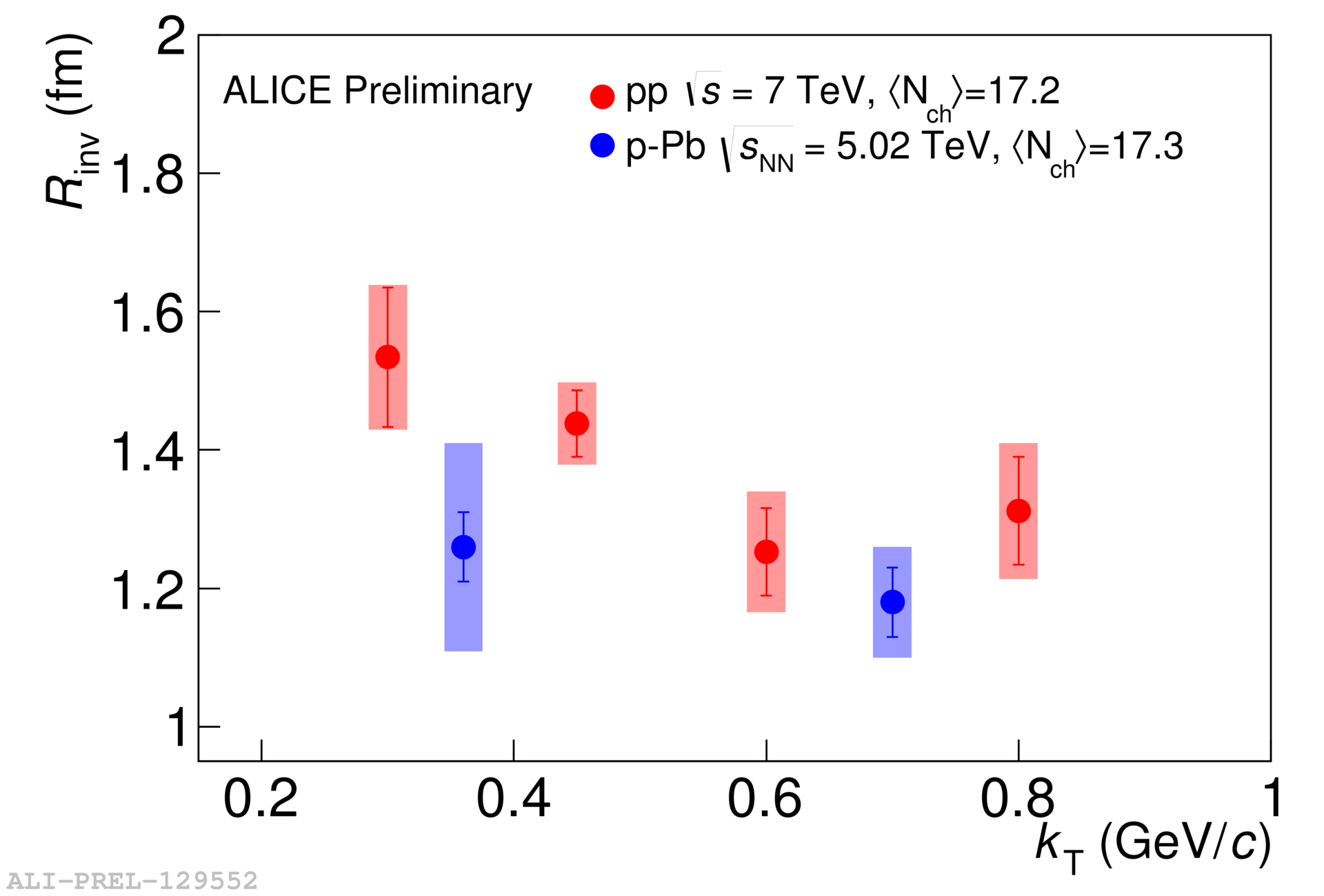}
\caption{Comparison of femtoscopic radii, as a function of pair transverse momentum $k_{\rm T}$,
obtained in pp \cite{Abelev:2012sq} (red circles) and p--Pb (blue circles) collisions.}
\label{fig:Rinv_kt_ncheqv}
\end{center}
\end{figure}

\begin{figure}%[h]
\centering
\subfigure[~]{\label{fig:Rinv_NchA}\includegraphics[width=0.44\textwidth]{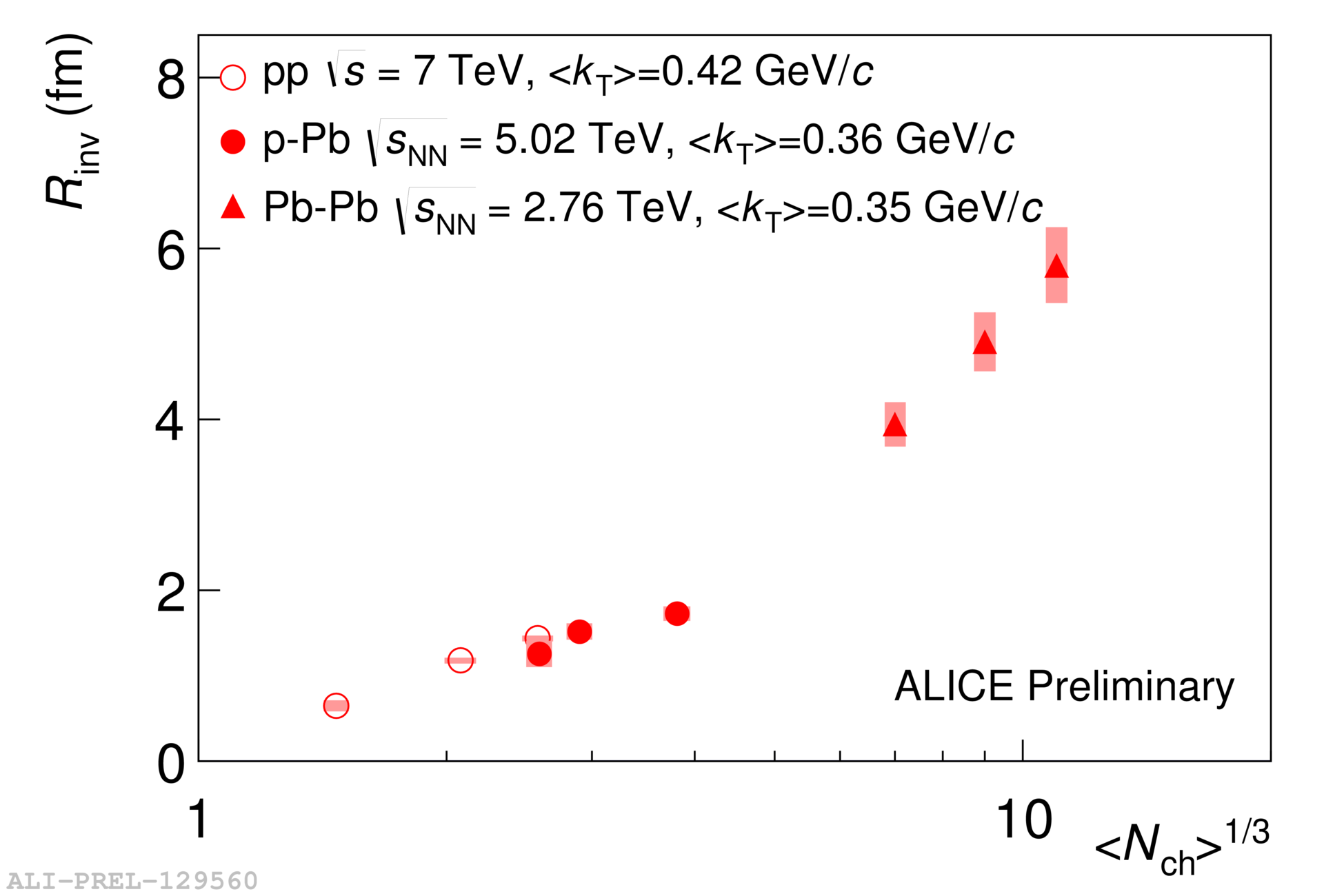}}~~~~~
\subfigure[~]{\label{fig:Rinv_NchB}\includegraphics[width=0.44\textwidth]{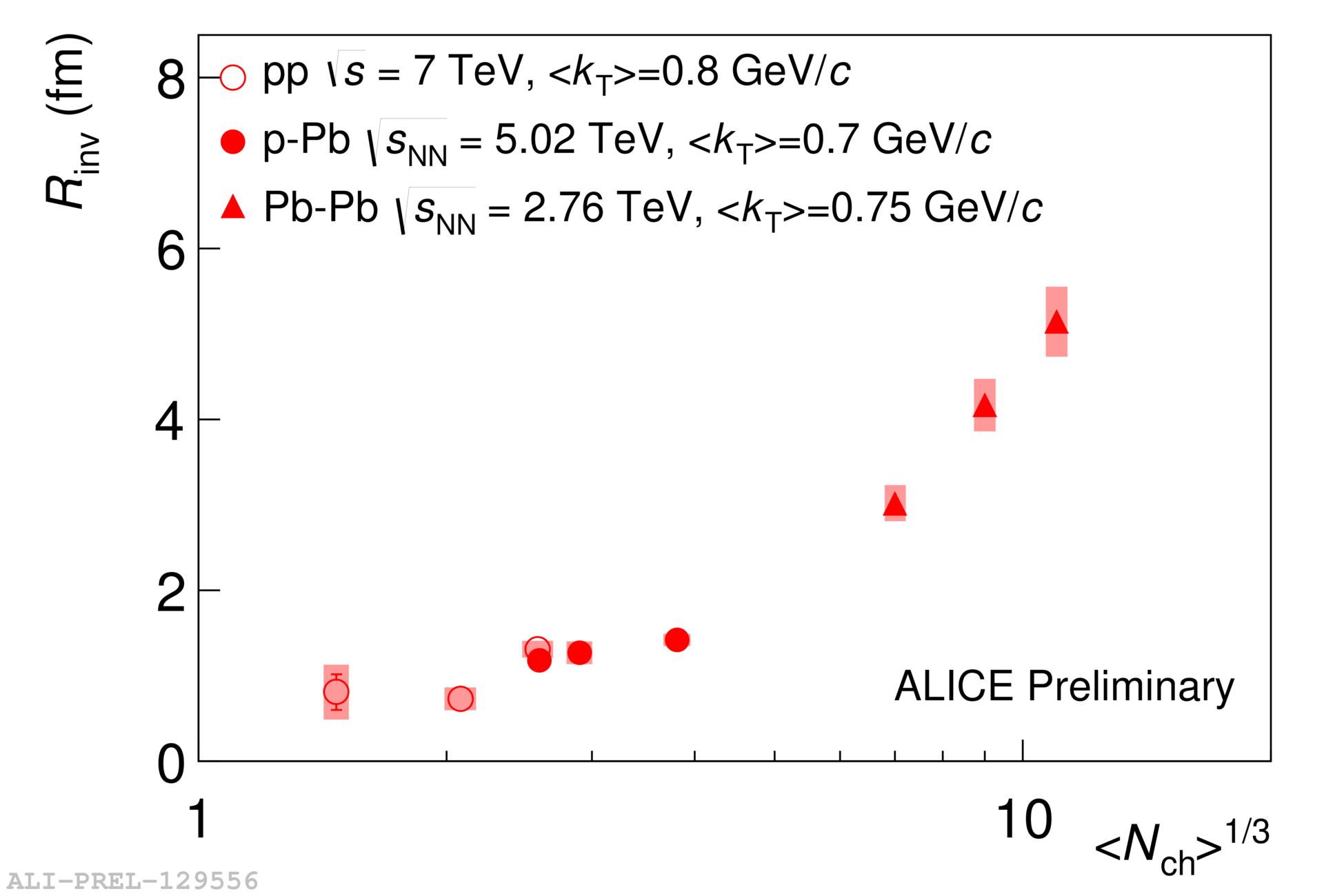}}
\caption{Comparison of femtoscopic radii, as a function of the measured charged-particle multiplicity density $N_{\rm ch}$, at low (a) and high $k_{\rm T}$ (b)
obtained in pp \cite{Abelev:2012sq} (empty circles), p--Pb (full circles) and Pb--Pb (full triangles) \cite{Adam:2015vja} collisions.}
\label{fig:Rinv_Nch}
\end{figure}

\begin{figure}%[h]
\centering
\subfigure[~]{\label{fig:Lam_NchA}\includegraphics[width=0.49\textwidth]{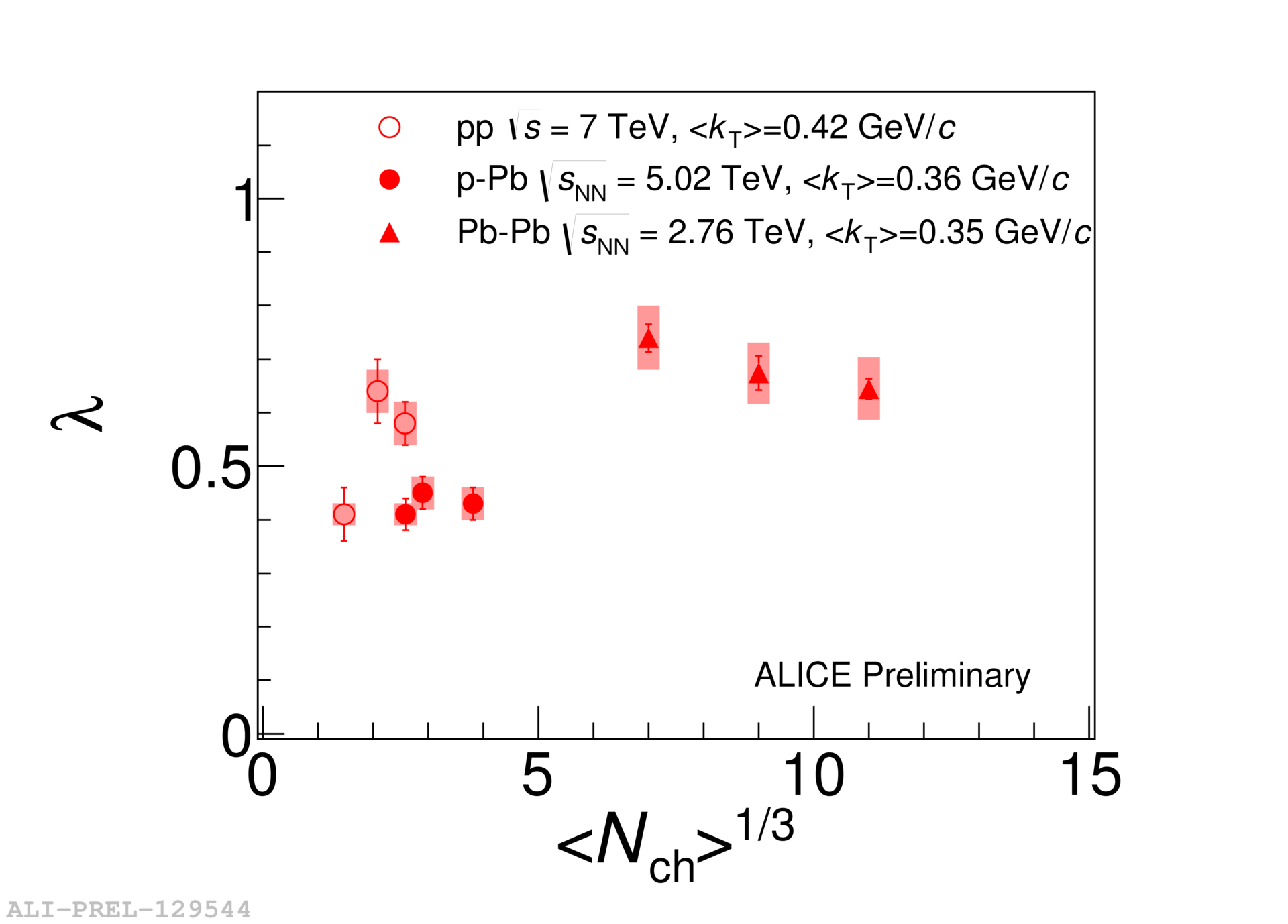}}
\subfigure[~]{\label{fig:Lam_NchB}\includegraphics[width=0.49\textwidth]{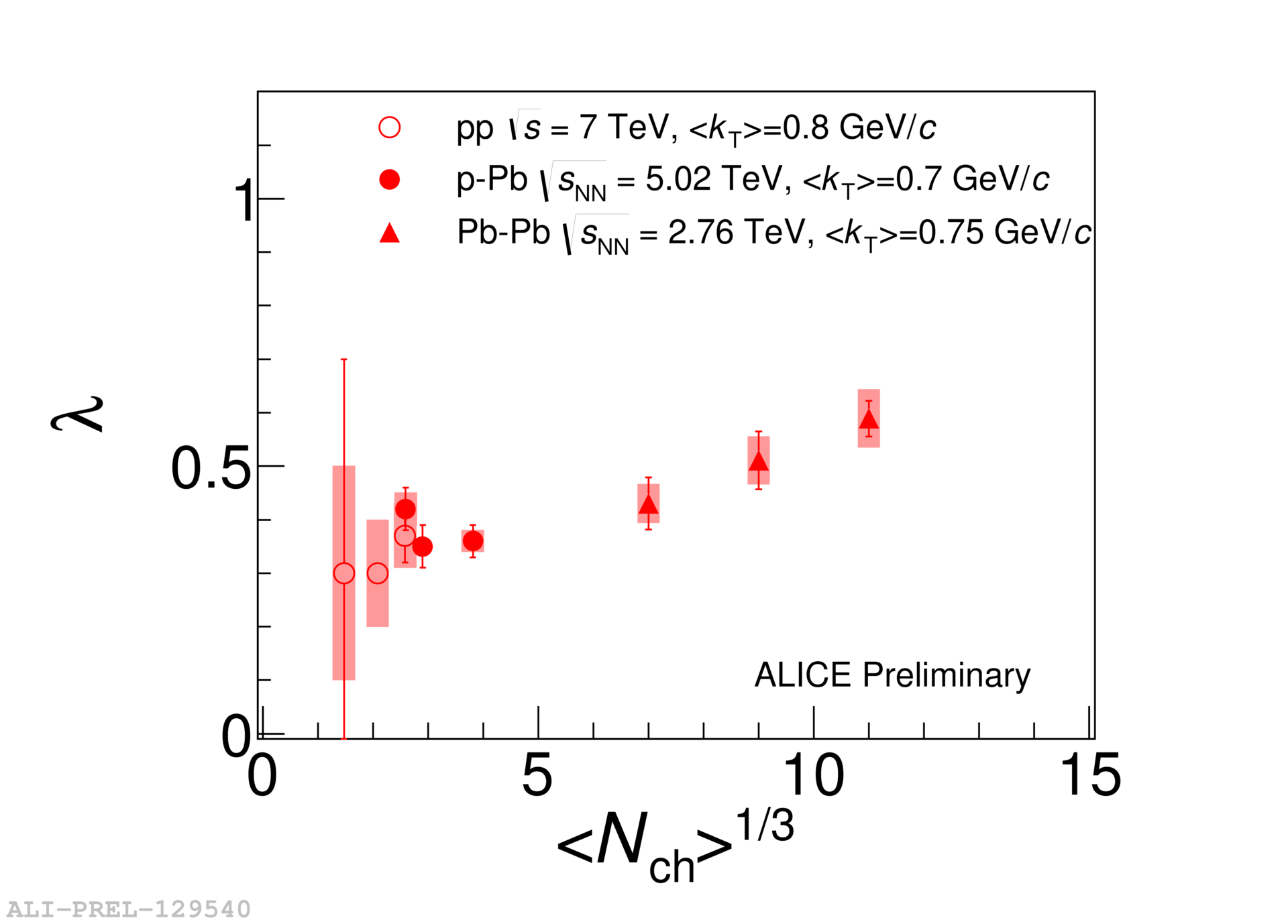}}
\caption{Comparison of correlation strengths $\lambda$, as a function of the measured charged-particle multiplicity density $N_{\rm ch}$, at low (a) and high (b) $k_{\rm T}$
obtained in pp \cite{Abelev:2012sq} (empty circles), p--Pb (full circles) and Pb--Pb (full triangles) \cite{Adam:2015vja} collisions.}
\label{fig:Lam_Nch}
\end{figure}

\begin{figure}[h]
\begin{center}
\includegraphics[width=0.5\textwidth]{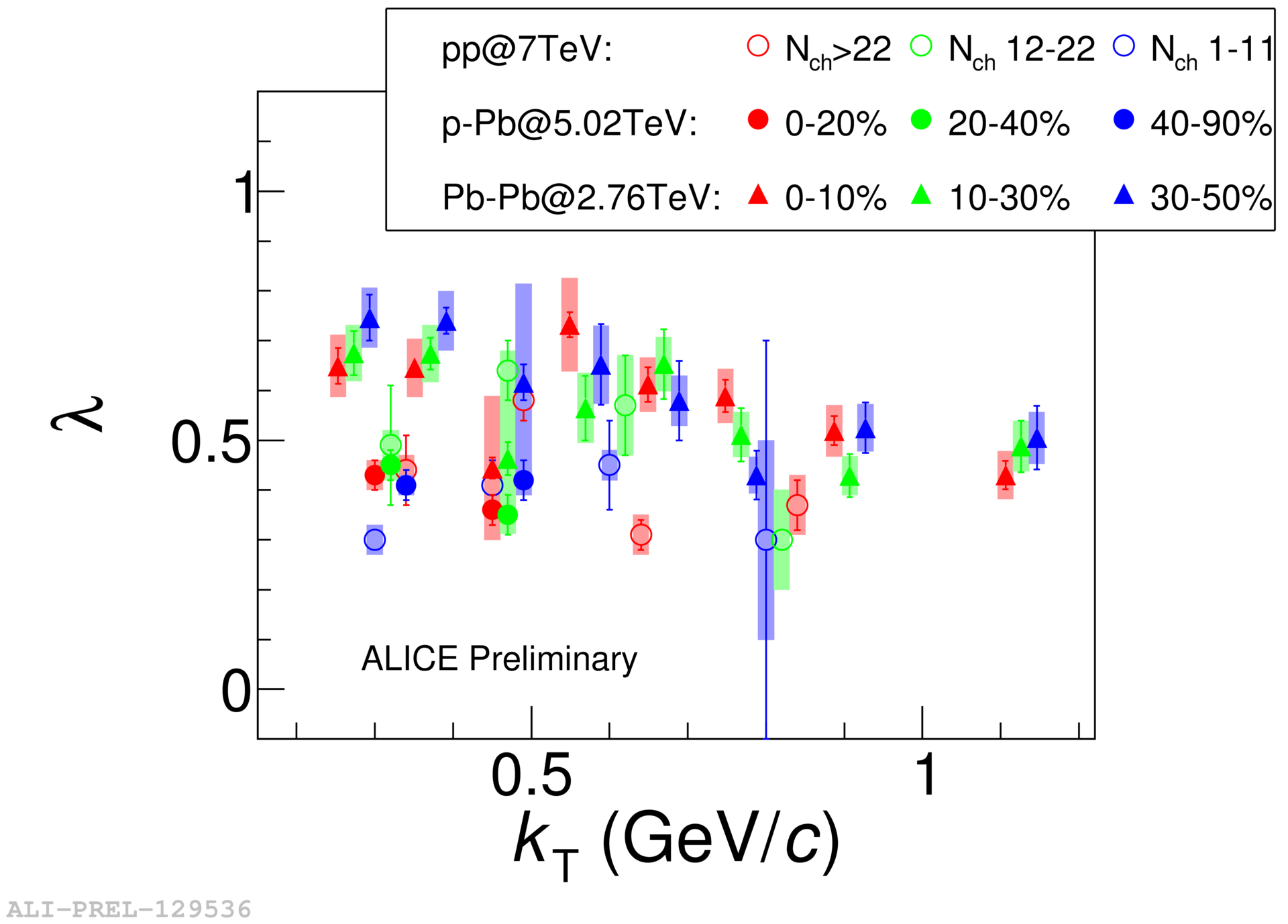}
\caption{The K$^{\pm}$K$^{\pm}$ correlation strengths $\lambda$ in pp \cite{Abelev:2012sq}, p--Pb and Pb--Pb \cite{Adam:2015vja} collisions versus pair transverse momentum $k_{\rm T}$ in all centrality and $k_{\rm T}$ bins.}
\label{fig:Lam_all}
\end{center}
\end{figure}

The extracted kaon $R_\mathrm{inv}$ and $\lambda$ are depicted in figures~\ref{fig:RL_sysA} and \ref{fig:RL_sysA}, respectively. %where statistical errors are shown by lines and systematic uncertainties -- by full rectangles.
In Fig.~\ref{fig:Rinv_kt_ncheqv}, the radii from pp collisions at $\sqrt{s}=7$~TeV \cite{Abelev:2012sq} and p--Pb collisions at $\sqrt{s_{\rm NN}}=5.02$~TeV at similar multiplicity are compared as a function of pair transverse momentum $k_{\rm T}$. The corresponding radii in Pb--Pb collisions at $\sqrt{s_{\rm NN}}=2.76$~TeV \cite{Adam:2015vja} are not shown since they cannot be extracted at comparable multiplicities due to a lack of data. The figure shows that at the same multiplicity, the radii in p--Pb collisions are equal to those in pp collisions within uncertainties. This observation differs from the results of the three-dimensional $\pi\pi$ \cite{Adam:2015pya} and one-dimensional three-pion cumulants \cite{Abelev:2014pja} analyses, where radii in pp collisions were obtained 5--15\% less than those in p--Pb collisions.
Figure \ref{fig:Rinv_Nch} compares femtoscopic radii, as a function of the measured charged-particle multiplicity density $<N_{\rm ch}>^{1/3}$, at low \ref{fig:Rinv_NchA} and high \ref{fig:Rinv_NchB} $k_{\rm T}$ extracted in pp \cite{Abelev:2012sq}, p--Pb and Pb--Pb \cite{Adam:2015vja} collisions. The obtained radii increase with $N_{\rm ch}$ and follow the multiplicity trend observed in pp collisions. The radii are equal in p--Pb and pp collisions at similar multiplicity within uncertainties. This result indicates the absence of strong collective expansion in p--Pb collisions at low multiplicities \cite{Bozek:2013df}.
As it is seen from the figure, the radii in p--Pb and Pb--Pb collisions cannot be compared at the same $N_{\rm ch}$. In order to make a conclusion as it was done in the pion correlation analyses \cite{Adam:2015pya,Abelev:2014pja} more experimental data are needed.
Figures \ref{fig:Lam_NchA} and \ref{fig:Lam_NchB} show the correlation strengths $\lambda$ in pp \cite{Abelev:2012sq}, p--Pb and Pb--Pb \cite{Adam:2015vja} collisions at low and high $k_{\rm T}$, respectively.
All $\lambda$ values are less than unity and vary in the range $0.3<\lambda<0.7$ due to the influence of long-lived resonances and a non-Gaussian shape of the kaon CF peak. It can be noticed from the figure that the correlation strength parameters in Pb--Pb collisions tend to be higher than those in pp and p--Pb collisions. That could point to a more Gaussian source created in Pb--Pb collisions.
Figure \ref{fig:Lam_all} compares correlation strengths $\lambda$ in pp \cite{Abelev:2012sq}, p--Pb and Pb--Pb collisions as a function of $k_{\rm T}$. These correlation strengths vary in the range $0.3<\lambda<0.7$. There is no noticeable $k_{\rm T}$ or centrality dependence observed.

\section{Conclusion}
In this work, one-dimensional identical charged kaon correlations have been obtained and analyzed for the first time in proton--nucleus collisions, that is in p--Pb at $\sqrt{s_{\rm NN}}=5.02$~TeV. The source size $R_{\rm inv}$ and correlation strength $\lambda$ have been extracted from a correlation function parametrized in terms of invariant pair relative momentum $q_{\mathrm{inv}}$. The obtained radii $R_{\rm inv}$ decrease with increasing pair transverse momentum $k_{\rm T}$ and with decreasing event multiplicity. This is similar to the behavior of pion radii in the three-dimensional two-pion correlation %\cite{Adam:2015pya}
and one-dimensional three-pion cumulant analyses%\cite{Abelev:2014pja}
. The kaon $R_{\rm inv}$ in p--Pb and pp collisions lie on the same multiplicity trend, whereas it is difficult to compare them with the results from Pb--Pb collisions because of a big multiplicity gap. The results disfavor models which incorporate substantially stronger collective expansion in p--Pb collisions as compared to pp collisions at similar multiplicity%\cite{Bozek:2013df}
. The correlation strength $\lambda$ varies from 0.3 to 0.7 for all collision systems, with no noticeable dependence on centrality or $k_{\rm T}$. The fact that the correlation strength in Pb--Pb collisions tends to be higher than in pp and p--Pb collisions could be an indication of a more Gaussian source created in Pb--Pb collisions. However, a strong conclusion is prevented due to large statistical uncertainties.